\documentclass[prl,twocolumn,amsmath,amssymb,floatfix,superscriptaddress]{revtex4-1}

\usepackage{graphicx}
\usepackage{amssymb}
\usepackage{amsmath}
\usepackage{color}
\usepackage{ wasysym }

\def\barray{\begin{array}}
\def\earray{\end{array}}
\def\be{\begin{equation}}
\def\ee{\end{equation}}
\def\ben{\begin{equation} \nonumber}
\def\een{\end{equation}}
\def\ban{\begin{eqnarray*}}
\def\ean{\end{eqnarray*}}
\def\ba{\begin{eqnarray}}
\def\ea{\end{eqnarray}}

\def\({\left(}
\def\){\right)}

\graphicspath{{./fig/}}

\begin{document}

\title{Reheating After Swampland Conjecture}

\author{Vahid.Kamali}
\affiliation{Department of Physics, McGill University, Montreal, QC, H3A 2T8,
Canada}
\affiliation{Department of Physics, Bu-Ali Sina (Avicenna) University, Hamedan 65178,
016016, Iran}
\affiliation{School of Physics,
Institute for Research in Fundamental Sciences (IPM),
19538-33511, Tehran, Iran}




\begin{abstract}
{ 
The evolution of the universe started from a hot and dense Big Bang point. Temperature fluctuation map of cosmic microwave background (CMB) radiation and initial seeds of large scale structures (LSS) are explained by an inflationary period in a very early time. Inflaton as quanta of inflation field is responsible for the accelerated expansion of the universe. Potentials of the self-interacting single field models are constrained by observational data as well as quantum gravity. Some forms of the potential are rolled out by data of Planck satellite and some of them by quantum gravity constraints. In the standard model of inflation or cold inflation firstly universe expands where the inflaton rolls the nearly flat part of the potential and in the second part, the universe reheats where the inflaton oscillates around the minimum of the potential which leads to thermalized radiation dominated universe. String theory as the best model of quantum gravity forbids the oscillation around the minimum of the potential during the thermalized epoch of the reheating. But in the warm model of inflation thermalization happens during the expansion of the universe where the inflaton rolls nearly steep potential and the universe will be radiation dominated without any separated reheating epoch.} 
\end{abstract} 
\maketitle 
 {\bf Motivation:} 
Big Bang (BB) model is originally introduced based on the expansion of the galaxies(clusters). This model has a famous contradiction with the background temperature of the CMB observational data which is known as horizon problem. This problem can be resolved by an inflation scenario which introduces an accelerated expansion period of the universe evolution in early time. The idea of inflation was proposed in the context of quantum field theory (QFT) \cite{Guth:1980zm, Linde:1981mu, Albrecht:1982wi}. 
Other more interesting features of the inflation are the explanation of the initial seed of LSS production and CMB temperature anisotropy which are studied in the context of cosmological perturbation theory \cite{Mukhanov:1990me}. The standard cold model of inflation \cite{Linde:1981mu}, in term of second-order phase transition, is described by two main parts, first part is the accelerated expansion of the universe which is presented by inflaton field, rolling slowly, and the second part is the reheating epoch as a connection between inflaton dominated era and radiation dominated era which is explained by oscillation of the inflaton around the minimum of the potential transferring its energy to mainly light standard particles \cite{Kofman:1994rk, Shtanov:1994ce}.

It was conjectured that the scalar field theory of inflation model should not obey the slow-roll conditions, this is swampland conjecture \cite{Obied:2018sgi,
Garg:2018reu,Agrawal:2018own,Garg:2018reu,Ooguri:2018wrx}. 
Some efforts have been done to resolve this contradiction \cite{Akrami:2018ylq,
Achucarro:2018vey,
Kehagias:2018uem,Denef:2018etk,Brahma:2018hrd,Lin:2018kjm,
Dimopoulos:2018upl,Ashoorioon:2018sqb,Wang:2018kly,Olguin-Tejo:2018pfq,Park:2018fuj,Lin:2018rnx,Schimmrigk:2018gch,Yi:2018dhl,Cheong:2018udx,Ibe:2018ffn,Blanco-Pillado:2018xyn,Andriot:2018mav,Kamali:2018ylz,Kinney:2018kew,Lin:2018edm,Bastero-Gil:2018yen,Corvilain:2018lgw,Scalisi:2018eaz,Seo:2018abc,Arciniega:2018tnn,Kallosh:2019axr,Kiritsis:2019wyk}. 
The conjecture of swampland motivates us to study the reheating epoch as a more problematic part of the early time cosmology. The main contradiction between the conjecture and effective field theory of early time cosmology is not slow-roll part of inflation but is reheating part which can be resolved in the context of the warm inflation model \cite{Berera:1995ie, Berera:1995wh, Berera:1996fm,Bastero-Gil:2016qru,Kamali:2019ppi}. 
The warm scenario of inflation is an alternative model of accelerated expansion of the universe where the thermalization happens during the slow-roll epoch and inflation era connects to radiation dominated era smoothly without any separated reheating epoch. This model with the suppressed tensor-to-scalar ratio is more compatible with observation data \cite{Akrami:2018odb} and solves the contradiction between swampland conjecture and the standard model of inflation both slow-roll and reheating parts.

{\bf The Swampland Conjecture:}
The theory of string suggests a vast of the landscape of vacua which are surrounded by maybe bigger swampland low-energy-looking-consistent semi-classical effective field theories (EFT) coupled to gravity. The EFTs are physically consistent with the quantum theory of gravity if:   
\begin{eqnarray}\label{c1}
\frac{\triangle\phi}{M_p}<c_1
\end{eqnarray}
where $\triangle\phi$ is excursion of the scalar fields in the field space, and:  
\begin{eqnarray}\label{c2}
\frac{\vert\nabla V(\phi)\vert}{V}>\frac{c_2}{M_p}\\
\nonumber
or\\
\nonumber
\frac{\min(\nabla_i\nabla_jV(\phi))}{V}\leq -\frac{c_3}{M_p^2},\\
\nonumber
\end{eqnarray}
where $V(\phi)$ is the potential of low-energy EFT, $c_i\sim\mathcal{O}(1)$ are universal constants and $\min(\nabla_i\nabla_jV(\phi)$ is defined in an orthonormal frame as Hessian eigenvalue minimum \cite{Obied:2018sgi, Agrawal:2018own}.
 The second case in Eq.(\ref{c2}) is refined de Sitter swampland conjecture \cite{Garg:2018reu, Ooguri:2018wrx}.  In this letter, we will examine EFT of reheating epoch, after the slow-roll expansion of the universe, in the context of the string swampland conjecture.

 {\bf Thermal field theory and phase transitions:}
Known or hypothetical early universe models of cosmology have been discussed by quantum field theories which are coupled to gravity. These theories are the low-energy limit of string theory models. In cosmological thermal systems with temperature $T$, which is comparable with the energy scales of the cosmological system or Hubble parameter ($T\sim H$), the thermodynamic potential instead of scalar field potential $V(\phi)$ is important:
\begin{eqnarray}
V(\varphi,T)=V(\varphi)+\frac{1}{24}m^2(\varphi)T^2-\frac{\pi^2}{90}T^4+Q.C,
\end{eqnarray} 
where $T>m$,~$m^2(\varphi)=\frac{d^2V}{d\varphi^2}$, Q.C is quantum corrections and $\varphi=\langle\phi\rangle$ is the expectation value of the scalar field in a thermal equilibrium as thermodynamic variable.

Using a toy but important symmetry breaking scalar field model with potential: 
\begin{eqnarray}
V(\varphi)=V_0-\frac{1}{2}\mu^2\varphi^2+\frac{1}{4}\lambda\varphi^4,
\end{eqnarray}
we can study phase transition which has a crucial role at early time cosmology. 
This potential without thermal corrections has two minima at $\varphi_{min}=\pm\frac{\mu}{\sqrt{\lambda}}$, where the thermodynamic system settles into one of them as broken phase and a maximum at $\varphi_{max}=0$.

\begin{figure}
\begin{center}
\includegraphics[scale=0.25]{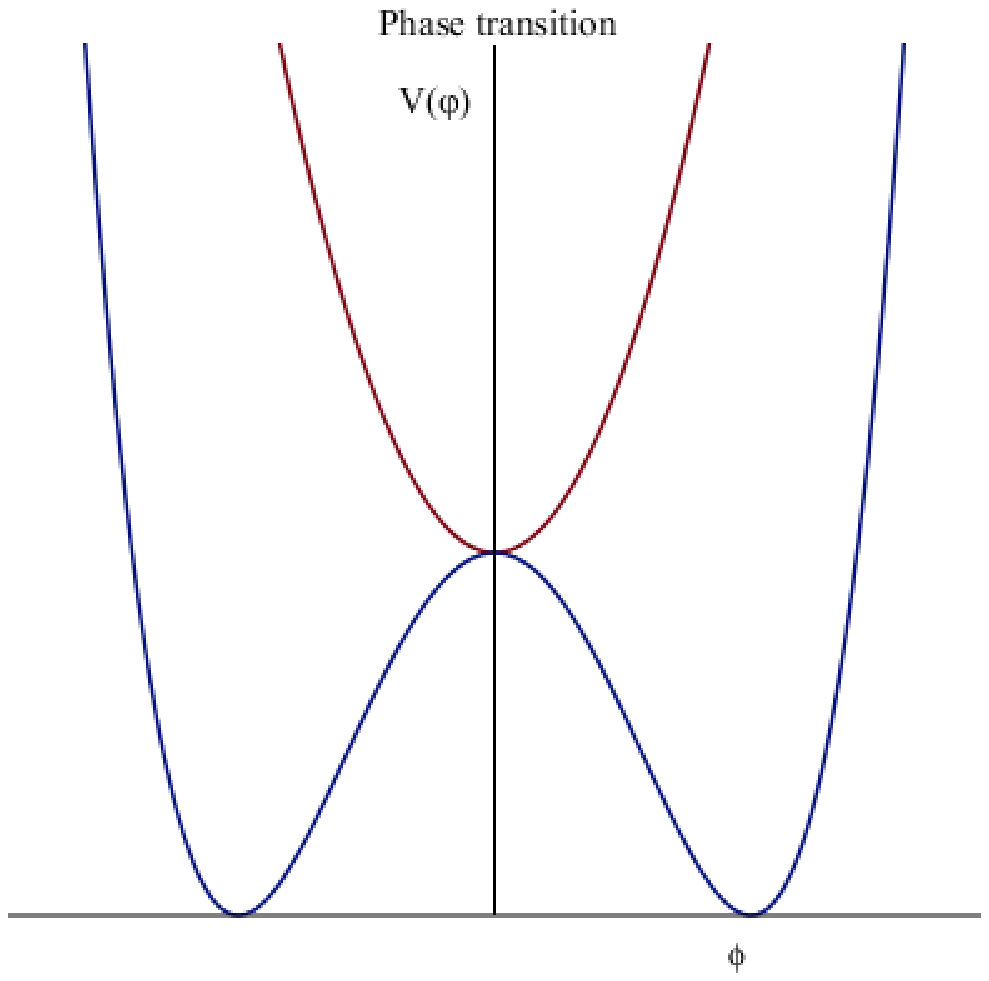}
\end{center}
\caption{The phase transition happens between symmetric case (red graph) and symmetry broken case (blue graph) during early time evolution of the universe.}
\label{Ph-Tr1}
\end{figure}

\begin{figure}
\begin{center}
\includegraphics[scale=0.25]{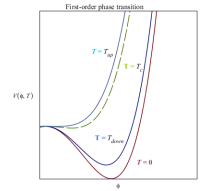}
\end{center}
\caption{Old inflation as a first order phase transition is in term of tunneling between false vacuum (local minimum at $\phi=0$) and true vacuum (global minimum at $\phi>0$). In this figure we can compare the symmetric phase with $T>T_c$  and broken phase with $T<T_c$. }
\label{Fi-Or1}
\end{figure}

\begin{figure}
\begin{center}
\includegraphics[scale=0.25]{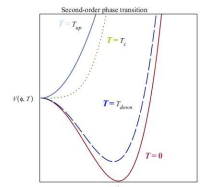}
\end{center}
\caption{ New inflation model is presented by the second-order phase transition between two cases: $T=T_{up}$ and $T=0$. Inflation is mainly explained by $T=0$ curve where at the first step inflaton rolls slowly down the nearly flat part of the potential and in the second step near the minimum of the potential oscillates heating the universe. At the reheating epoch, thermal correction modified the potential as long-dashed blue curve $T_{down}$ which is in contrast with swampland conjecture.}
\label{Se-Or1}
\end{figure}

When the thermal corrections become important the shape of the potential as a function of $\varphi$ is modified. At temperature $T_{up}$ which is higher than critical temperature $T_c$ the thermodynamic system settles into a new symmetric phase with the new minimum $\phi=0$ (Fig.\ref{Ph-Tr1}). This procedure is a phase transition between broken and symmetric phases which can be first or second order. $T_c$ is the critical temperature of the thermal system at the phase transition point.
In the first-order phase transition when the temperature changes between $T_{low}$ and $T_{up}$ (where $T_{low}<T_c<T_{up}$) there are two local minima, with a barrier between them, in the shape of modified potential(see Fig.\ref{Fi-Or1}). 
If our system is firstly in temperature $T_{up}$ and the temperature falls below critical temperature $T_c$, the system has to return some time in metastable state $\varphi=0$  because of the potential barrier which separates false vacuum $\varphi=0$ and true vacuum $\varphi=\frac{\mu}{\sqrt{\lambda}}$.
At critical temperature $T=T_c$ the values of potential are similar for two local minima Fig.\ref{Fi-Or1}. The phase transition between $T_{up}$ and $T_{low}$ case can be done by quantum tunneling of the potential barrier which leads to bubble nucleation in the broken phase.  
In second-order phase transition case when the system has temperature $T_{up}$ the shape of modified potential has just a minimum and after phase transition to broken phase with temperature, $T_{low}$ the shape of the potential has a local minimum and a local maximum Fig.\ref{Se-Or1}. Therefore there is no barrier and also no metastable state.
The idea of old inflation \cite{Guth:1980zm,Guth:1979bh} was introduced in term of scalar field theory which obeys the first-order phase transition.
This idea has a "graceful exit" problem. Transition to the thermal universe is the main problem of this model. In term of old inflation concepts during the first-order phase transition, the produced bubbles expand and cross each other.
The gradient of the scalar field at the boundary of the crossed bubbles leads to a kind of kinetic energy density  which could produce the amount of energy which is needed for the universe reheating.
But the problem of this idea is that the bubbles could not cross each other because of the expansion of the universe. The idea of old inflation in term of first-order phase transition could not explain the thermal epoch after inflation and the temperature of nucleosynthesis epoch. 
Therefore the old model of inflation was not expected as a workable model in cosmology. The idea of new inflation was introduced soon after the old one in term of second-order phase transition \cite{Linde:1981mu,Albrecht:1982wi}.

In this scenario the universe expands during short time slow-roll epoch when the kinetic energy density is smaller than potential energy density $V(\phi),$ and perturbation modes of the scalar field cross the horizon and freeze out. These perturbations lead to curvature perturbations in the context of general relativity.
The early perturbations are initial seeds of the large scale structure and temperature fluctuations of cosmic microwave background (CMB). The results of the observational data analysis for the fluctuations of CMB temperature can be used to constrain models of inflation \cite{Akrami:2018odb}. 
After accelerated expansion or slow-roll epoch of the universe evolution where the inflaton field, rolls slowly the nearly flat part of potential it goes to the minimum of the potential Fig.\ref{Se-Or1}. Oscillation of inflaton around the minimum of the potential with the energy exchange to other mainly light fields, during nearly matter-like era $\rho\propto a^{-3}$, leads to the reheated universe.
       
{\bf Reheating:}
In reheating epoch the accelerated expansion of the universe terminates and the kinetic part of energy density is comparable with potential energy. Reheating has three steps \cite{Dolgov:1982th,Abbott:1982hn,Traschen:1990sw,Dolgov:1989us,Kofman:1994rk}, most of the inflaton energy density transfer to bosonic particles during non-perturbative broad parametric resonance or preheating epoch at first step.
The decay of bosonic particles to standard particles is the second step and finally, thermalization is the third step. The particle creations during the reheating epoch are usually studied by interacting two fields potential, for example:
\begin{eqnarray}\label{p1}
V(\phi,\chi)=\frac{1}{2}m_{\phi}^2\phi^2+\frac{1}{2}m_{\chi}^2\chi^2+\sigma\phi\chi^2+h^2\phi^2\chi^2+k\chi^4
\end{eqnarray}
where the inflaton field interacts with other fields in the model which is generally presented by $\chi$. Non-thermal phase transitions idea before standard particle creation with the symmetry breaking potentials
\begin{eqnarray}\label{p2}
V(\phi,\chi)=\frac{\lambda}{4}(\phi^2-\phi_0^2)^2+\frac{1}{2}g^2\phi^2\chi^2,~~~~~~\\
\nonumber
V(\phi,\chi)=\frac{\lambda}{4}\phi^4+\frac{\alpha}{4}(\chi^2-\frac{M^2}{\alpha})^2+\frac{1}{2}g^2\phi^2\chi^2,
\end{eqnarray}  
can be used to explain topological defects and cosmic strings problem before thermalized era  \cite{Kofman:1995fi}. 
Potentials (\ref{p1}) and (\ref{p2}) are explicitly in contrast with the second part of second swampland conjecture (de Sitter conjecture) which is about the condition of the potential form of the EFT.\\

\begin{figure}
\begin{center}
\includegraphics[scale=0.25]{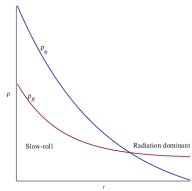}
\end{center}
\caption{During slow-roll warm inflation the radiation energy density is comparable with the scalar field energy density but smaller than it. At the end of the slow-roll part the radiation energy density smoothly dominates on the inflaton energy density.   }
\label{En-De1}
\end{figure}

{\bf Discussion:}
The contradiction between potentials (\ref{p1},\ref{p2}) and the second part of de Sitter  conjecture may not be the case if the potential provides the first condition of de Sitter conjecture. The main problem between reheating and de Sitter conjecture is thermalization part. At the thermalization step, for all models of inflation, the form of potential near the minimum is approximately quadratic, on the other hand, there is a thermal bath with a temperature $T$. Considering thermal part of reheating we need a thermal effective field theory with the potential (see the blue long-dashed curve in Fig.\ref{Se-Or1}):
\begin{eqnarray}
V(\varphi,T)\simeq\frac{1}{2}m_{\varphi}^2\varphi^2+\frac{1}{24}m^2(\varphi)T^2-\frac{\pi^2}{90}T^4+QC,
\end{eqnarray}  
{There is another alternative for the thermal modification of the potential if we consider the fermionic contributions to the effective potential \cite{Cline:1996mga, Bastero-Gil:2016qru}
\begin{eqnarray}
V_T\simeq-\frac{7\pi^2}{180}+\frac{m^2 T^2}{12}+\frac{m^4}{16\pi^2}[\ln(\frac{\mu^2}{T^2})-c_f]
\end{eqnarray}
where $m$ is the mass of the fermion, $\mu$ is the renormalization scale and $c_f=2.635$. All this modifications can generally alleviate the de Sitter constriant as we can see in Fig.\ref{Se-Or1}. The main point is that the modified potential is obviously in contrast with the first part of de Sitter conjecture at the minimum of the potential (see long-dashed blue curve in Fig.\ref{Se-Or1}). 
The potential also could not obey the second part of the conjecture which is originally considered near the maximum of the potential. These mean that the thermal EFT of reheating is suffered from swampland string theory conjecture.} 
The solution of this concern is warm inflation model \cite{Berera:1995ie, Berera:1996fm, 
Bastero-Gil:2016qru, Kamali:2019ppi}.
In the context of warm inflation, inflaton field interacts with the light fields during the slow-roll epoch and the inflation era connects to radiation dominated era smoothly (see Fig.\ref{En-De1}). 
The universe heats up during the slow-roll epoch of warm inflation and there is no need to the separated reheating epoch. On the other hand, there were some discussions about slow-roll part of warm inflation which cover the swampland conjecture \cite{Das:2018hqy,Motaharfar:2018zyb,Das:2018rpg,Bastero-Gil:2018yen}. { In warm inflation model the background evolution of the inflaton is modified  by the interaction of inflaton with light fields during the sow-roll part:
\begin{eqnarray}
\ddot{\varphi}+3H(1+Q)\dot{\varphi}+\frac{dV}{d\varphi}=0
\end{eqnarray} 
This new dissipation parameter $Q$ has an important effect on the swampland discussion where the Hubble slow-roll parameter is modified as:
\begin{eqnarray}
\epsilon_{H}=-\frac{\dot{H}}{H^2}=\frac{1}{1+Q}\frac{1}{2}M_{pl}(\frac{V'}{V})^2 <1.
\end{eqnarray}
This  slow-roll condition can be achieved by a steep potential, this is the de Sitter swampland condition. On the other hand having a steep potential leads to a small excursion of the scalar field  which is the distance swampland conjecture \cite{Motaharfar:2018zyb,Das:2018rpg,Bastero-Gil:2018yen}. }

 All solutions of the contradictions between string theory and (cold)inflation theory that have been proposed in the literature \cite{Akrami:2018ylq,Achucarro:2018vey,Kehagias:2018uem,
Denef:2018etk,Brahma:2018hrd,Lin:2018kjm,
Dimopoulos:2018upl,Ashoorioon:2018sqb,Wang:2018kly,Olguin-Tejo:2018pfq,Park:2018fuj,Lin:2018rnx,Schimmrigk:2018gch,Yi:2018dhl,Cheong:2018udx,Ibe:2018ffn,Blanco-Pillado:2018xyn,Andriot:2018mav,Kamali:2018ylz,Kinney:2018kew,Lin:2018edm,Bastero-Gil:2018yen,Corvilain:2018lgw,Scalisi:2018eaz,Seo:2018abc,Arciniega:2018tnn,Kallosh:2019axr,Kiritsis:2019wyk}
may helpful for the slow-roll part but are unable to explain the contradiction between reheating epoch and string swampland conjecture, as we have discussed in this letter.
The main conclusion of our work is introducing warm inflation as a solution to swampland conjecture, slow-roll and reheating problems. 
We emphasize that the cold model of inflation is rolled out by quantum gravity constraints on reheating epoch but warm inflation without any reheating epoch is acceptable.     
 
{\bf Acknowledgements:}
I want to thank Cumrun Vafa, Andrei Linde,  Robert Brandenberger, Shahin Sheikh-Jabbar, Amjad Ashoorioon, Ali-Akbar  Abolhasani, Rudnei Ramos, Meysam Motaharfar, Ahmad Mehrabi, Morteza Salehi, Mohammad Malekjani and Pourya Asari for some valuable discussions. The research at McGill has been supported by a NSERC Discovery Grant to Robert Brandenberger and also been supported in
part by the McGill Space Institute. 
 \bibliographystyle{apsrev4-1}
  \bibliography{ref}

\end{document}